\documentclass[9pt]{article}

\usepackage[USenglish]{babel} 
\usepackage[T1]{fontenc}

\usepackage{lmodern} 
\usepackage{setspace}

\usepackage{amsmath}
\usepackage{amsthm}
\usepackage{amsfonts}
\usepackage{fancyhdr}
\usepackage{graphics}
\usepackage{epsfig}
\usepackage{graphicx}
\usepackage{textcomp}
\usepackage{amsmath}
\usepackage{amsfonts}
\usepackage{amssymb}
\usepackage{mathrsfs}
\usepackage{amsthm}
\usepackage{enumerate} 
\usepackage{epstopdf} 
\usepackage{float}
\usepackage{subcaption}
\usepackage{cleveref}

\usepackage{placeins} 
\usepackage{flafter}  
\usepackage{verbatim} 
\numberwithin{equation}{section}

\begin{document}
\begin{titlepage}
\title{Acoustic Superradiance from a Bose-Einstein Condensate Vortex with a Self-Consistent Background Density Profile}
\author{ Bet\"ul Demirkaya\footnote{bdemirkaya@ku.edu.tr}, Tekin Dereli\footnote{tdereli@ku.edu.tr}, Kaan G\"uven\footnote{kguven@ku.edu.tr},  \\ {\small Department of Physics, Ko\c{c} University, 34450 Sar{\i}yer, \.{I}stanbul, Turkey }}
 
\date{\today}
\maketitle
\begin{abstract}
\noindent

The axisymmetric acoustic perturbations in the velocity potential of a Bose-Einstein condensate in the presence of a single vortex behave like minimally coupled massless scalar fields propagating in a curved (1+1) dimensional Lorentzian space-time, governed by the Klein-Gordon wave equation. Thus far, the amplified scattering of these perturbations from the vortex, as a manifestation of the acoustic superradiance, has been investigated with a constant background density. This paper goes beyond by employing a self-consistent condensate density profile that is obtained by solving the Gross-Pitaevskii equation for an unbound BEC. Consequently, the loci of the event horizon and the ergosphere of the acoustic black hole are modified according to the radially varying speed of sound. The superradiance is investigated both for transient features in the time-domain and for spectral features in the frequency domain. In particular, an effective energy-potential function defined in the spectral formulation correlates with the existence and the frequency dependence of the acoustic superradiance. The numerical results indicate that the constant background density approximation underestimates the maximum superradiance and the frequency at which this maximum occurs.

\end{abstract}
\vskip 2cm
\end{titlepage}

\maketitle			
\clearpage 
\section{Introduction}
Unruh  introduced the idea of condensed-matter analogies of gravitational systems by showing that acoustic perturbations in the velocity potential of a locally irrotational, barotropic, inviscid Newtonian fluid behave exactly as a minimally coupled massless scalar fields propagating in a curved(3 + 1) dimensional Lorentzian spacetime \cite{Unruh1981}. Since then, different condensed-matter and optical systems have been studied to demonstrate the analogies of various cosmic-scale gravitational phenomena down at the laboratory scale.  \cite{barcelo2011analogue}.
In the last few years,
experimental realizations of black- and white-horizons were reported in water channels \cite{PhysRevLett.106.021302}, atomic Bose-Einstein condensates (BECs) \cite{PhysRevLett.105.240401}, nonlinear optical fibers and silica glass 
illuminated by strong laser pulses, and also in nonlinear
optical systems \cite{PhysRevA.86.063821},\cite{philbin2008fiber}. 
In 2014, Acoustic black hole in in a
needle-shaped BEC of 87Rb  is achieved and recently spontaneous Hawking radiation, stimulated by quantum vacuum fluctuations, emanating from an analogue black hole in an atomic Bose-Einstein
condensate is reported \cite{steinhauer2014observation},\cite{steinhauer2016observation}.

  One of the striking features of rotating atomic BECs is the formation of vortices \cite{CARLSON1996183}. The occurrence of  vortices in superfluids has been the focus of fundamental theoretical and experimental works \cite{fetter2001vortices}, \cite{macher2009black}, \cite{garay2001sonic}. Scattering process from a single vortex, the superradiance phenomena, has been analyzed theoretically and numerically for constant density approximation in BECs as well as in classical fluid \cite{basak2003superresonance}, \cite{torres2017rotational}.

Motivated by the recent experimental progress in BEC systems, we
present here a consolidating study of the temporal and spectral features
of the scattering process from a BEC vortex with non-constant background density with a modified version of the Visser's draining bathtub model(DBT) \cite{barcelo2001analogue}. 
Constant density, therefore constant speed of sound approximation provide certain freedom in choosing the fluid velocity and the scaling factor.   
However, for a single unbound vortex in BEC, rotational velocity is naturally quantized [cite] and the radial velocity is shaped by the continuity equation with the defined density profile, which  shapes the speed of sound throughout the vortex. Since velocities and density are linked together, we loose the freedom to increase the vortex fluid without changing the  profiles for  density and the speed of sound. However, for a stable single vortex  these assumptions are physically more plausible.

In this paper we analyzed the two main assumptions constant and non-constant density assumptions, and compare the result within the superradiance calculations.  Single unbound vortex density is calculated via the Gross-Pitaevskii equation which gives a density profile that sets the non-constant speed of sound  with a natural scaling factor, healing length. The time domain solutions
are obtained by solving the Klein-Gordon equation for the propagation of
acoustic waves by implementing the
numerical techniques described mainly in Ref.s \cite{num1},\cite{cherubini2005},\cite{cherubini2006}, whereas the spectral analysis of the superradiance is conducted by asymptotic solutions of the waves at the event horizon and the
ergosphere. In addition wave equation in frequency domain is solved, numerically, to calculate the reflection coefficient after the coordinate transformation in order to avoid the singularity, event horizon.

\section{The vortex state of an unbound BEC condensate}

The single vortex state in the condensate of weakly interacting bosons is described by the Gross-Pitaevskii (GP) equation,
\begin{equation}
i\hbar\frac{\partial\Psi}{\partial t}=\left(-\frac{\hbar^2}{2m}\nabla^2+ V_{ext} +U_0 \left| \Psi \right|^2\right)\Psi(r,t)
\label{GP}
\end{equation}
where $V_{ext}$ is the external potential and $V_H=\frac{4a\pi\hbar^2}{m}\left| \Psi \right|^2 \equiv U_0\left| \Psi \right|^2$ is the repulsive Hartree potential for binary interactions among atoms, that is characterized by the scattering length, $a$, $V_H$ prevents the collapse instability. 

For an unbounded condensate (i.e $V_{ext} = 0$),  on dimensional grounds, the balance between the kinetic energy and $V_H$ implies a correlation(healing) length
\begin{equation}
  \xi^2=\frac{\hbar^2}{2\rho_{\infty}mU_0}= \frac{1}{8\pi \rho_{\infty} a }\\ 
\label{healing}
 \end{equation}
 $\rho_{\infty}$ donates the  bulk density of the uniform condensate. The healing length describes the distance at which the condensate wave function  tends to its bulk form in the presence of localized perturbations \cite{fetter2001vortices},\cite{pethick2002bose}.

The stationary solutions take the form $\Psi(r,t)=\psi(r)e^{-i\mu t/\hbar}$ with a background density $\rho(r)= \left|\psi(r) \right|^2$ and chemical potential $\mu=\frac{\hbar^2}{2m\xi^2}$. For a vortex state, the radial part has a axisymmetric formal solution  in polar coordinates 
\begin{equation}
\psi(r,\theta) = f_q(r) e^{iq\theta}, 
\label{eq:wave}
\end{equation}
where the subscript $q$ donates the  winding number.  Substituting Eq.\ref{eq:wave} into Eq.\ref{GP} yields
\begin{equation}
\mu \psi(r)=-\frac{\hbar^2}{2m} \frac{1}{r}\frac{\partial }{\partial r} \left(r \frac{\partial f_q}{\partial r}\right)+
\frac{\hbar^2}{2mr^2}f_q + U_0 f_q^3.
\label{GPform}
\end{equation}
Using dimensionless radial coordinate $\tilde{r}=r/\xi $ and the scaled function $ \chi_q= f_q / \sqrt{\rho_{\infty}}$ the equation takes the form 
\begin{equation}
\frac{1}{\tilde{r}} \frac{d}{d\tilde{r}}\left(\tilde{r}\frac{d\chi_q}{d\tilde{r}}\right) + \frac{\chi_q}{\tilde{r^2}} + \chi^3_q - \chi_q=0.
\label{eq:onevortdens}
\end{equation}
 Subject to the boundary conditions
\begin{align}
\chi_q(0)=0 && \chi_q(\infty)=1.
\end{align}

\begin{figure}[H]
	\centering
		\includegraphics[width=0.8\textwidth]{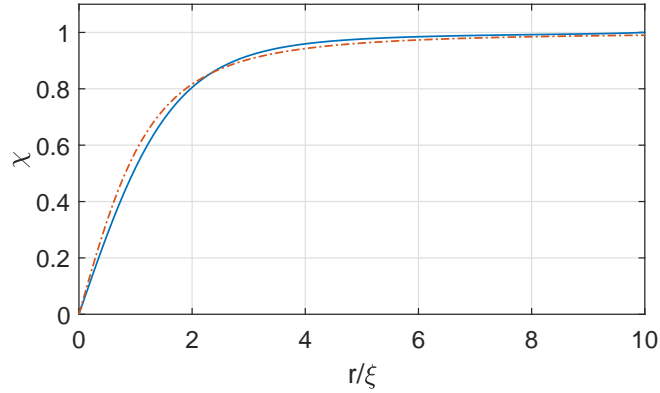}
	\caption{The radial wave function $\chi(r/\xi)$ obtained by numerical solution of the stationary GP
equation for a straight vortex line with the approximate form given in Eq.\ref{eq:density profile} shown by a dashed line.}
	\label{fig:density}
\end{figure}

 The vortices with  respective winding numbers $q = \pm 1$ are  found to be topologically stable but that those with larger values of $\left|q\right|$ are unstable and should decay into $\left|q\right|=1$ single-charge vortices \cite{desyatnikov2005optical}, \cite{zhang2007numerical}. Thus, Eq.\ref{eq:onevortdens}  is solved for $q=1$ only. Although a closed analytical form of solution is not available, an approximate functional  form is given by \cite{pethick2002bose},\cite{slatyer2005}

\begin{equation}
\chi_1= \sqrt{\frac{\tilde{r}^2}{2 + \tilde{r}^2 }},
\label{eq:density profile}
\end{equation}
 which provides a good approximation to the numerical solution as shown in Fig.\ref{fig:density}. In the next section, the approximate form will be employed to proceed with the theoretical formulation, but the full solution will be used for obtaining the numerical results.

\subsection{Dynamics of the vortex}

The definition of velocity in BEC is irrotational, which is a typical characteristic of superfluids \cite{fetter2001vortices}. However, in the presence of a vortex, the velocity is singular at the vortex center and the circulation around a contour enclosing the vortex becomes nonzero. Hence, the velocity acquires a tangential component, described by a winding number
	\begin{eqnarray}  
 \Theta = \oint \vec \upsilon \cdot \vec dl = 2\pi q \frac{\hbar}{m}, && q= \pm 1, \pm 2...\\
\upsilon_{\hat{\theta} }=  \frac{ \Theta}{2\pi r } \hat{\theta}.
\label{eqthetav}
\end{eqnarray}
In order to satisfy the continuity equation in cylindrical coordinates $ \frac{{\partial \rho }}{{\partial t}} + \nabla  \cdot (\rho \vec \upsilon ) = 0 $, the radial component of the velocity becomes
	\begin{equation}
	\nabla  \cdot (\rho_0 \vec \upsilon ) = 0 \Rightarrow \upsilon _{\hat r}= -\frac{A}{\rho(r) r}
	\label{eq:radv}
	\end{equation}
Hence, the draining bathtub model described in Ref.s \cite{dolan2012resonances},\cite{visser1998acoustic} has to be modified accordingly:
\begin{equation}
 \vec{ \upsilon}=\frac{-A}{r}\hat{r}+\frac{B}{r}\hat{\phi} \rightarrow  \vec{ \upsilon}=\frac{-A}{\rho(r)r}\hat{r}+\frac{B}{r}\hat{\phi}. \\ 
\label{fluidvel}
 \end{equation}
The parameters $A$ and $B$ are described in the  next section.

\section{Solution to Klein-Gordon equation in time domain}

With the radially varying background density profile, the GP equation can now be solved for first order fluctuations $\psi=  \psi_0 + \psi_1 =\sqrt{\rho} e^{i\Phi}$, against the background $\psi_0 = \sqrt{\rho_0 }e^{i\Phi_0}$. This yields a set of two equations. The first one is the continuity equation
 \begin{equation}
\label{cont}
            \frac{{\partial \rho }}{{\partial t}} + \nabla  \cdot (\rho \vec \upsilon ) = 0
            \end{equation}
and the so-called Euler equation with an additional term(first term on right-hand side of Eq.\ref{euler}) which is called the quantum pressure.
            \begin{equation}
						\label{euler}
            {\partial_t\vec \upsilon  + (\vec\upsilon  \cdot \vec\nabla )\vec\upsilon }  =  \frac{\hbar^2}{2m^2}\nabla\left(\frac{1}{\sqrt{\rho}}\nabla^2\sqrt{\rho}\right)-\frac{\nabla V_{ext}}{m}-\frac{\nabla(U_0\rho)}{m}.
            \end{equation}
Eq.s \ref{cont} and \ref{euler} are expressed in terms of the velocity of the condensate defined as
\begin{equation}
\vec \upsilon=(\hbar/m)\nabla \Phi.
\label{eq:velocity}
\end{equation}
The dynamics of small fluctuations can be formulated by linearizing the equations against the background: $\rho= \rho_0 + \rho_1$, $\Phi= \Phi_0 + \Phi_1$. This yields the equation system

\begin{equation}
            \frac{{\partial \rho_0}}{{\partial t}} + \nabla  \cdot (\rho_0 \vec \upsilon ) = 0
						\label{1eq}
            \end{equation}
	\begin{equation}
 \partial_t \Phi_0 = \frac{\left|\nabla \Phi_0\right|^2}{2m^2} + \frac{V_{ext}}{m} + \frac{U_0\rho_0}{m} - \frac{\hbar^2}{2m}\frac{\nabla^2\sqrt{\rho_0}}{\sqrt{\rho_0}}
\label{eqpsi0}
\end{equation}

 \begin{equation}
            \frac{{\partial \rho_1 }}{{\partial t}} + \frac{\hbar}{m}\nabla  \cdot (\rho_0 \nabla \Phi_1) + \nabla  \cdot (\rho_1 \vec\upsilon)= 0,
						\label{rho1}
  \end{equation}
	
  \begin{equation}
    \partial_t \Phi_1 =-\upsilon \cdot \nabla \Phi_1 - \frac{U_0}{\hbar}\rho_1 + \frac{\hbar^2}{2m}D_2 \rho_1,
		\label{phi1}
 \end{equation}
						where $D_2$ is 
\begin{equation}
D_2=\frac{1}{2\sqrt{\rho_0}}\nabla^2 \frac{\rho_1}{\sqrt{\rho_0}}-\frac{\rho_1}{2\rho^{3/2}_{0}}\nabla^2\sqrt{\rho_0}.
\label{qunapres}
\end{equation}
Note that Eq.s \ref{1eq}-\ref{eqpsi0} indicate that the background satisfies itself the GP equation whereas Eq.s \ref{rho1}-\ref{phi1} govern the fluctuation dynamics. 
For small fluctuations, $D_2$ is negligible in comparison to other terms \cite{pethick2002bose}. This establishes the hydrodynamic (quasi-classical) approximation, under which Eq.s \ref{rho1} and \ref{phi1} are combined to yield a single equation for the phase fluctuations

 \begin{equation}
\frac{\partial}{\partial t}\left[\frac{\rho_0}{c^2}\left(\frac{\partial \Phi_1}{\partial t} + \vec \upsilon \cdot \nabla \Phi_1\right]\right) - \nabla \cdot \left(\rho_0 \nabla \Phi_1\right) + \nabla \cdot \left[\frac{\rho_0}{c^2}\left(\frac{\partial \Phi_1}{\partial t} + \vec \upsilon \cdot \nabla \Phi_1\right)\right]=0.
\label{kleinbec}
 \end{equation}
Now, the  background density profile can be introduced by the approximate functional form for a single vortex state through Eq.\ref{eq:density profile}  with $\rho(r)= \left|\psi(r,\theta) \right|^2$ resulting in
\begin{equation}
\rho_0(r)=\rho_{\infty}\frac{(r-r_0)^2}{(r-r_0)^2+2\xi^2}
\label{dprof}
\end{equation}
where $r_0$ is the vortex center, $\xi$ is  the healing length and $\rho_{\infty}= \frac{1}{8\pi\xi^2a}$ is the bulk density far away from the vortex. The propagation speed of sound in a BEC  is given by $c=\sqrt{U_0\rho/m}$ which becomes a radially varying function near the vortex:
\begin{equation}
c(r)= c_{\infty}\sqrt{\frac{(r-r_0)^2}{(r-r_0)^2+2\xi^2}},
\label{speedofsound}
\end{equation}
where $c_{\infty}=\frac{\hbar}{m}\sqrt{4\pi a \rho_{\infty}}$ denotes the bulk value for sound speed. Taking $r_0=0$ and using cylindrical coordinates, Eq.\ref{kleinbec} can be written in explicit form as 
 \begin{align}
    &\left[ { \frac{{{\partial ^2}}}{{\partial {t^2}}} + 2\upsilon_r\frac{{{\partial ^2}}}{{\partial t\partial r}} + \frac{{2\upsilon_{\theta} }}{r}\frac{{{\partial ^2}}}{{\partial t\partial \theta }} + \left( -{c^2  + \upsilon^2_r} \right)\frac{{{\partial ^2}}}{{\partial {r^2}}} + \frac{{2\upsilon_{\theta}\upsilon_r }}{{r}}}\frac{{{\partial ^2}}}{{\partial r\partial \theta }}\right. \nonumber\\
  &\left. { + \left( -{\frac{c^2}{r^2} + \frac{\upsilon^2_{\theta}}{r^2}} \right) \frac{{{\partial ^2}}}{{\partial {\theta ^2}}} + \frac{1}{r}\frac{\partial(r\upsilon_r)}{\partial r}\frac{\partial }{{\partial t}} + \left({-\frac{\upsilon_{\theta}\upsilon_{r}}{r^2} +\frac{\upsilon_{\theta}}{r}\frac{\partial\upsilon_r}{\partial r}}\right)\frac{\partial }{{\partial \theta}}} \right. \nonumber\\
	&\left.{ + \left( {-\frac{c^2}{r} -\frac{c^2}{\rho} \frac{\partial \rho(r)}{\partial r} + \frac{\upsilon^2}{r} + 2\upsilon_r\frac{\partial \upsilon_r}{\partial r} + \upsilon_{\theta}\frac{\partial \upsilon_{\theta}}{\partial r}	}\right)\frac{\partial}{\partial r} - c^2\frac{\partial^2}{\partial z^2}}\right]\Phi_1  = 0.\label{kleingordon}
  \end{align}

In the following, the numerical techniques described in Ref.s \cite{num1},\cite{cherubini2005} are being implemented (the $\Phi_1$ of the present formulation corresponds to $\Psi$ in these references). Two conjugate fields are introduced: 

\begin{align}
\Gamma = \frac{\partial {\Phi}}{\partial x^i} && {\Pi}=-\frac{1}{c}\left(\frac{\partial {\Phi}}{\partial t}  - \beta^i {\Gamma}_i\right),
\label{conjugate}
\end{align}
where $x_i= (r, \theta ,z)$ denote the cylindrical coordinates and $\beta^i = (-\upsilon_r, - \upsilon_{\theta}/r, 0)$ are teh velocity components. The fields are introduced formally as
\begin{align}
{\Phi} = \phi_1 (t,r)e^{im\phi}e^{ikz}&& {\Pi} = \pi_1 (t,r)e^{im\phi}e^{ikz} && {\Gamma} = \gamma_1 (t,r)e^{im\phi}e^{ikz}
\end{align}
with $(k,m)$ denoting the axial and azimuthal wave numbers. In this study, we take  $k=0$, i.e. translational symmetry along the $z$ axis. These functional forms satisfy a set of first order coupled partial differential equations

\begin{equation}
\frac{\partial\phi_1}{\partial t} = -c\pi_1 - \upsilon_r\gamma_1 -\frac{im\upsilon_\theta}{r}\phi_1
\label{set1}
\end{equation}
	
\begin{align}
\frac{\partial\gamma_1}{\partial t} = &- \frac{\partial c}{\partial r}\pi_1 - c\frac{\partial \pi_1}{\partial r} -\frac{\partial \upsilon_r}{\partial r} \gamma_1- \upsilon_r\frac{\partial \gamma_1}{\partial r} \nonumber \\ 
&- \frac{\partial \upsilon_\theta}{\partial r} \frac{im}{r}\phi_1 + \frac{im\upsilon_\theta}{r^2}\phi_1 - \frac{im\upsilon_\theta}{r}\frac{\partial\phi_1 }{\partial r}
\label{set2}
\end{align}
	
	\begin{align}
\frac{\partial\pi_1}{\partial t} = &\pi_1\left(-\frac{\partial c}{\partial r}\frac{\upsilon_r}{c} - \frac{im\upsilon_\theta}{r} -\frac{1}{r}\frac{\partial(r\upsilon_r)}{\partial r} \right) -\upsilon_r\frac{\partial\pi_1}{\partial r } \nonumber \\ 
&+\frac{\gamma_1}{c}\left(-\frac{c^2}{r} - \frac{c^2}{\rho}\frac{\partial \rho}{\partial r}   + \frac{\upsilon_\theta^2 }{r} + \upsilon_\theta \frac{\partial \upsilon_\theta}{\partial r} \right) -c\frac{\partial\gamma_1 }{\partial r}\nonumber \\ 
&+\frac{\phi_1}{c}\left(\frac{m^2c^2 }{r^2} + c^2k^2- \frac{im\upsilon_r}{r}\frac{\partial\upsilon_\theta }{\partial r} -\frac{im\upsilon_r\upsilon_\theta }{r^2}      \right)
\label{set3}
\end{align} 
	
In section \ref{ASintime}, the superradiance will be investigated through the propagation of these conjugate fields by solving these equations numerically.
	
\subsection{The Event Horizon and the Ergosphere}

\begin{figure}[H]
	\centering
		\includegraphics[width=0.8\textwidth]{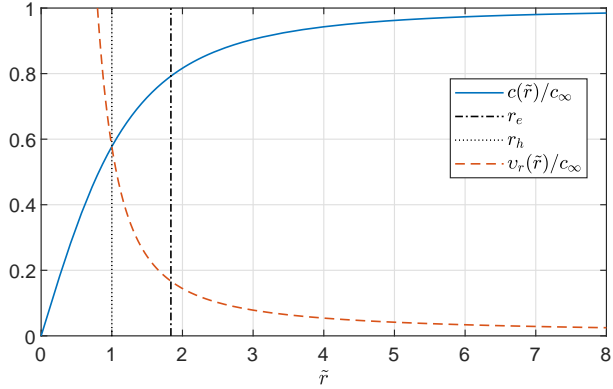}
	\caption{Representation of the event horizon, $r_h$, ergoregion, $r_e$, the speed of sound, $c(r)$ and the radial velocity of the fluid, $\upsilon_r(r)$ near vortex.}
	\label{velocities}
\end{figure}

The vortex state defines a curved 1+1 space-time where the line element reads
\begin{equation}
ds^2=\frac{\rho}{c}\left[-cdt^2 + (dr-\upsilon_{r}dt)^2 + (rd\theta - \upsilon_{\theta}dt)^2 \right].
\end{equation}
A new coordinate system is introduced through the following transformations that minimize the number of off-diagonal elements in the metric, thereby revealing the event horizon and the ergosphere.
	 \begin{align}
    &dt=dt^{*}-\upsilon_{{r}}/\left(c^2 - \upsilon^2_{{r}}\right)dr,\\
		& d\theta=d\theta^{*} - \upsilon_{{r}}\upsilon_{{\theta}}/\left(r\left(c^2 - \upsilon^2_{{r}}\right)\right)dr,  \\
		&r=r^{*},     z=z^{*},
\end{align}
For simplicity, the new coordinates are renamed as the old ones from now on. The line element takes the form
\begin{equation}
ds^2=\left[\left(\upsilon^2 -c^2\right)dt^2 + \frac{c^2}{c^2-\upsilon_{{r}}^2}dr^2 + r^2d\theta^2 + 2\upsilon_{{\theta}}rdtd\phi \right].
\label{line}
\end{equation}
In general relativity, the radius of the ergosphere, $r_e$, for a Kerr-type black hole is defined through the vanishing of the coefficient of $dt^2$, whereas the event horizon, $r_h$, is determined by the singularity of the metric. These conditions read respectively as
\begin{equation}
\upsilon^2-c^2=0 \Rightarrow \frac{A}{\rho_{\infty}c_{\infty}}-\frac{r_h^4}{\left(r_h^2+2\xi^2\right)^{3/2}}=0
\label{eq:evhorizon}
\end{equation}

\begin{equation}
\upsilon^2_{{r}}-c^2=0 \Rightarrow  \frac{A^2\left(r_e^2+2\xi^2\right)^2}{\rho^2_{\infty}r^6_e} + \frac{B^2}{r_e^2}-\frac{c_{\infty}^2r_e^2}{r_e^2+2\xi^2}=0
\label{eq:ergoregion}
\end{equation}
The coefficients $A$ and $B$ can be chosen to set the event horizon at unit healing length $r_h=\xi$:
\begin{align}
A=\frac{\xi \rho_\infty c_\infty}{3^{3/2}} && B=\frac{\hbar}{m}.
\end{align}
Note that B follows from Eq.\ref{eqthetav}. As a practical example, we employ the parameters of a BEC of Rb-87 atoms taken from Ref.s\cite{dalfovo1999theory},\cite{anderson1995observation}: 
\begin{align}
a=5.77 nm && m=1.44 *10^{-25} kg && \rho_{\infty} =10^{21} m^{-3}  
\end{align}
which yields
\begin{align}
c_\infty = 6.2*10^{-3} m/s && \xi=83 nm
\end{align}
By substituting these values into Eq.\ref{eq:ergoregion}, the resulting polynomial of degree 8 in $r_e$ can be solved numerically, which has only one positive real root: $r_e=1.84\xi$. Figure \ref{velocities} shows the event horizon, the ergosphere along with the variation of the speed of sound and the radial velocity of the condensate fluid near the vortex.

\section{Acoustic superradiance in the time-domain}
\label{ASintime}

We adopt the numerical method dubbed as the "excision technique" to solve the Eq.s \cref{set1,set2,set3}, which provides a numerically feasible way to deal with the radial range beyond the event horizon, which is physically inaccessible. We refer the reader to Ref.\cite{cherubini2005,excision,demirkaya2018analog} for the details. In this work, we employ Matlab's PDE solver toolbox with an FDTD solver. The computational radial domain is set as  $0.5< \tilde{r} < 110$ and the propagation is computed in the range  $0 < \tilde{t} < 110$, where $\tilde{r}=r/\xi$, $\tilde{t}= t c_{\infty}/\xi $ are dimensionless coordinates.For the outer computational boundary, one has to implement absorbing boundary conditions or simply ignore it by completing the simulation before the outgoing wave reaches to the outer boundary. The inner computational boundary is set beyond the event horizon. The aforementioned excision technique introduces additional constraint equations at the event horizon that have to be monitored to ensure that no perturbation propagates from beyond the event horizon into the physically relevant part of the computational domain.

The incident perturbative wave is chosen to be
\begin{equation}
\psi_1(0,r)=Nexp\left[-(r-r_0+c(r)t)^2/b^2 - i\omega(r-r_0 +c(r)t)/c(r)\right].
\label{initial}
\end{equation}
This defines a cylindrically imploding Gaussian wave, initially centered at $r_0$, of width $b$. In the numerical calculations $r_0=50\xi$ and $b=10\xi$ are used. The initial values of  $\pi_1$ and $\gamma_1$ are calculated by Eq.\ref{conjugate}. It is worth to mention that, the density profile  given in Eq.\ref{eq:density profile} is based inherently on the scaling by the healing length, $\xi$. The velocity is scaled by $c_{\infty}$,

 \begin{equation}
\vec{ \tilde{\upsilon}} = - \frac{1}{3\sqrt{3}}\left(\frac{1}{\tilde{r}} + \frac{2}{\tilde{r}^3}\right)\hat{r}+ \frac{\sqrt{2}}{\tilde{r}}\hat{\theta} 
\label{eq:radvel}
\end{equation}
The energy of the perturbations is given by\begin{equation}
E(t)=\int \partial^3r \frac{1}{2}M\rho\vec{\upsilon}^2=(\hbar^2/2M)\int^{2 \pi}_{0}d \phi \int^{H}_{0}dz \int^{r_{max}}_{1} \rho(r)(\nabla \Phi)^2 r dr.
\label{eq:energy}
\end{equation}

\begin{figure}[H]
	\centering
		\includegraphics[width=0.6\textwidth]{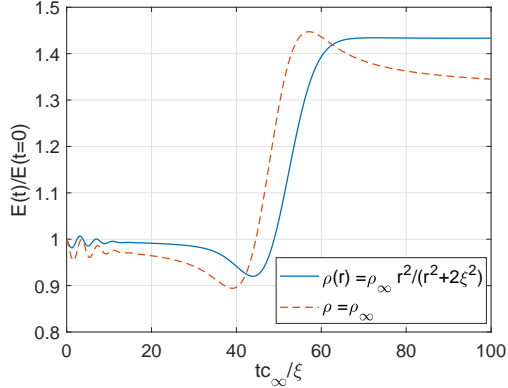}
	\caption{Time evaluation of the energy gain of the wave packet for superradiant case for constant and non-constant density profiles. Wave parameters used are  $r0=50\xi$ and $b=10\xi$ with $\omega=0.8c_\infty/\xi$}
	\label{consnoncons}
\end{figure}

Fig.\ref{consnoncons} shows the temporal change of the relative energy calculated by the constant background density approximation and by the present formulation. For comparison, the event horizon is set at $r_h=\xi$, same initial perturbative wave is applied in both cases. Nevertheless, the ergosphere and the radial speed differ, which delays the arrival of the incident perturbation to the event horizon for the non-constant density as seen in the Figure.  For the constant background density case, the amplification exhibits an overshooting transient behavior, whereas in the case of the non-constant background density, the amplification saturates monotonically to its final value. Overall, the non-constant background denisty case predicts a slightly higher ($\sim 8\%$) amplification.

\begin{figure}[H]
	\centering
		\includegraphics[width=0.6\textwidth]{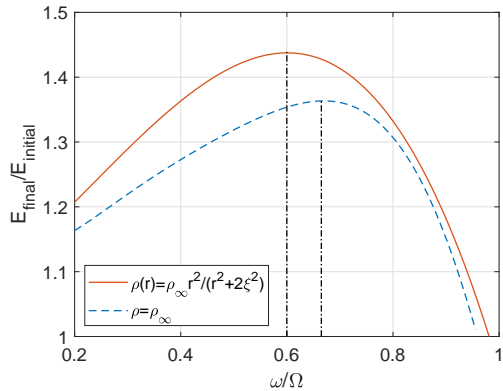}
	\caption{Energy gain difference between constant density and non-constant density approach for $\tilde{\Omega}/5<\tilde{\omega}<\tilde{\Omega}$. Parameters given in Fig.\ref{consnoncons} }
	\label{energycomp}
\end{figure}

In Figure \ref{energycomp} the spectra of the relative energy for the constant and non-constant background density cases are plotted respectively. Since $\vec{\upsilon_{\theta}}=\vec{\Omega} \times \vec{r} $, the angular frequency of the vortex is equal to
\begin{equation}
\Omega= \frac{B}{r_h^2} = \sqrt{2}c_\infty/\xi
\label{eq:Omega}
\end{equation}
A subtle but important difference is that, in the constant background density case, $\Omega$ appears as an independent and unlimited parameter, whose integer multiples, $m\Omega$, define vortex rotational speed. Of course, $m=1$ is known to describe a stable vortex. In contrast, the non-constant self-consistent density profile dictates the radial velocity and hence the value of $\Omega$ is "built-in" with a particular value that cannot be changed arbitrarily, unless the density profile is modified. 

The frequency range of the incident perturbative wave is taken as $0.2<\omega/\Omega<1$ since the superradiance is expected for $\omega<m\Omega$, and $m=1$ for stable vortices. The comparison with constant and non-constant density profiles indicate that in the latter case, the maximum superradiance is obtained at a slightly lower frequency.

\section{The spectral analysis of superradiance}

We begin by writing the Klein-Gordon equation (Eq.\ref{kleingordon}) with a formal solution of  the form:
\begin{equation}
 \Phi_1= R(r)e^{i(m\theta-\omega t)} 
\label{eq:wavefunc}
\end{equation}

\begin{equation}
\frac{\partial^2 R(r)}{\partial^2 r} + P(r) \frac{\partial R(r)}{\partial r} + Q(r)R(r)=0,
\label{eq:1}
\end{equation}
where 
\begin{equation}
P(r)= \left(-2i\omega \upsilon_r + \frac{2im\upsilon_r \upsilon_{\theta}}{r} - \frac{\upsilon^2_{\theta} + \upsilon^2_r-c^2}{r} - \frac{c^2}{\rho}\frac{\partial \rho}{\partial r} + \frac{\partial}{\partial r}\left(\upsilon^2_r+ \frac{\upsilon^2_\theta}{2} \right) \right)\frac{1}{\upsilon^2_r - c^2},
\label{eq:P}
\end{equation}
\begin{equation}
Q(r)=\left(-\omega^2 + \frac{2m\omega \upsilon_{\theta}}{r} - m^2\left(\frac{-c^2+\upsilon^2_{\theta}}{r^2}\right) -i\omega\left(\frac{1}{r}\frac{\partial(r\upsilon_r)}{\partial r}\right) - \frac{im\upsilon_{\theta}\upsilon_r}{r^2} + \frac{im\upsilon_{\theta}}{r}\frac{\partial \upsilon_r}{\partial r}\right)\frac{1}{\upsilon^2_r - c^2},
\label{eq:Q}
\end{equation}

According to the coefficient of $dr^2$ in the line element (Eq.\ref{line}), we define tortoise coordinate transformation $r\rightarrow r_*$, which maps the radial range $r\in(r_h,\infty)$ unto $r_*\in(-\infty,\infty)$
\begin{equation}
\Delta = \frac{d r_*}{dr} = \left(1-\frac{\upsilon^2_r}{c^2}\right)^{-1}
\label{eq:tort}
\end{equation}
Next, the radial part of the solution is expressed in a product form $R(r)= H(r_*)Z(r)$, where the functions respectively satisfy the following differential equations
\begin{equation}
\frac{\partial Z(r)}{\partial r} + M(r)Z(r)=0,
\label{eq:Z}
\end{equation}
\begin{equation}
\frac{\partial^2 H(r_*)}{\partial r_*^2} + V(r) H(r_*)=0;
\label{eq:H}
\end{equation}
The solution of Eq.\ref{eq:Z} is given by
Eq.\ref{eq:Z} gives 
\begin{align}
Z(r)&= C exp\left(-\int M(r) dr \right)\\
&=C \left(\frac{r^2 + 2\xi^2}{r^3}\right)^{1/2}exp\left(i\int h(r)dr\right)
\end{align}
where 
\begin{equation}
h(r)= \frac{\upsilon_r\left(\upsilon_{\theta} - r\omega\right)}{r\left(\upsilon^2_r - c^2\right)},
\label{eq:h}
\end{equation}
C is an arbitrary constant. We note that all velocity terms in Eq.\ref{eq:h} are functions of the radial coordinate.
Equation \ref{eq:H} is in the form of a time-independent Schrodinger equation with
\begin{equation}
V(r)=\frac{1}{\Delta^2}\left(\frac{\partial M}{\partial r} + M^2 + PM +Q\right)
\label{eq:V}
\end{equation}
\begin{equation}
M(r)=-\left(\frac{P}{2} + \frac{1}{2\Delta}\frac{\partial \Delta }{\partial r}\right)
\label{eq:M}
\end{equation}
Where the $V(r)$ reads explicitly
\begin{align}
V(r)=&\left(\frac{\omega}{c} - \frac{m\upsilon_\theta}{rc}\right)^2  + \frac{\left(\upsilon^2_r -c^2\right)\upsilon^2_r\left(5r^4 + 76r^2\xi^2 + 180\xi^4\right)}{4c^4r^2(r^2 + 2\xi^2)^2} \nonumber \\
&+ \frac{\left(\upsilon^2_r -c^2\right)\left(4m^2\left(r^2 + 2\xi^2\right)^2-r^4 - 20r^2\xi^2 + 12\xi^4\right)}{4c^2r^2\left(r^2 + 2\xi^2\right)^2}
\label{Vr} 
\end{align}

The first term is clearly a function of the energy of the incident perturbative wave while the last two terms depend on $r$ only. We note that $\upsilon_r(r)$ and $c(r)$ can be expressed in terms of the event horizon (through Eq.\ref{eq:evhorizon}) and the background density profile. Indeed, one can express $V(r)$ in dimensionless coordinates as

\begin{equation}
V(\tilde{r},\tilde{\omega})=E(\tilde{r},\tilde{\omega}) + V_1(\tilde{r})
\label{eq:}
\end{equation}

\begin{equation}
E(\tilde{r},\tilde{\omega}) = \frac{1}{\xi^2}\left(\frac{\tilde{r}^2+2}{\tilde{r}^2}\left(\tilde{\omega}-\frac{m\tilde{\Omega}}{\tilde{r}^2}\right)^2\right)
\label{eq:}
\end{equation}
\begin{align}
V_1(\tilde{r})=&-\left(\left(- 27\tilde{r}^8 + \tilde{r}^6 + 6\tilde{r}^4 + 12\tilde{r}^2 + 8\right)\left(2768\tilde{r}^2 + 2032\tilde{r}^4+ 696\tilde{r}^6 + 430\tilde{r}^8 \right.\right.\nonumber \\
&\left.\left. - 535\tilde{r}^{10} - 27\tilde{r}^{12} + 108m^2\tilde{r}^8\left(\tilde{r}^2 + 2\right) + 1440\right)\right)/(2916\xi^2\tilde{r}^{18}(\tilde{r}^2 + 2)^2)
\label{eq:V_1}
\end{align}

\begin{figure}[H]
    \centering
    \begin{subfigure}{0.4\textwidth}
        \includegraphics[width=\textwidth]{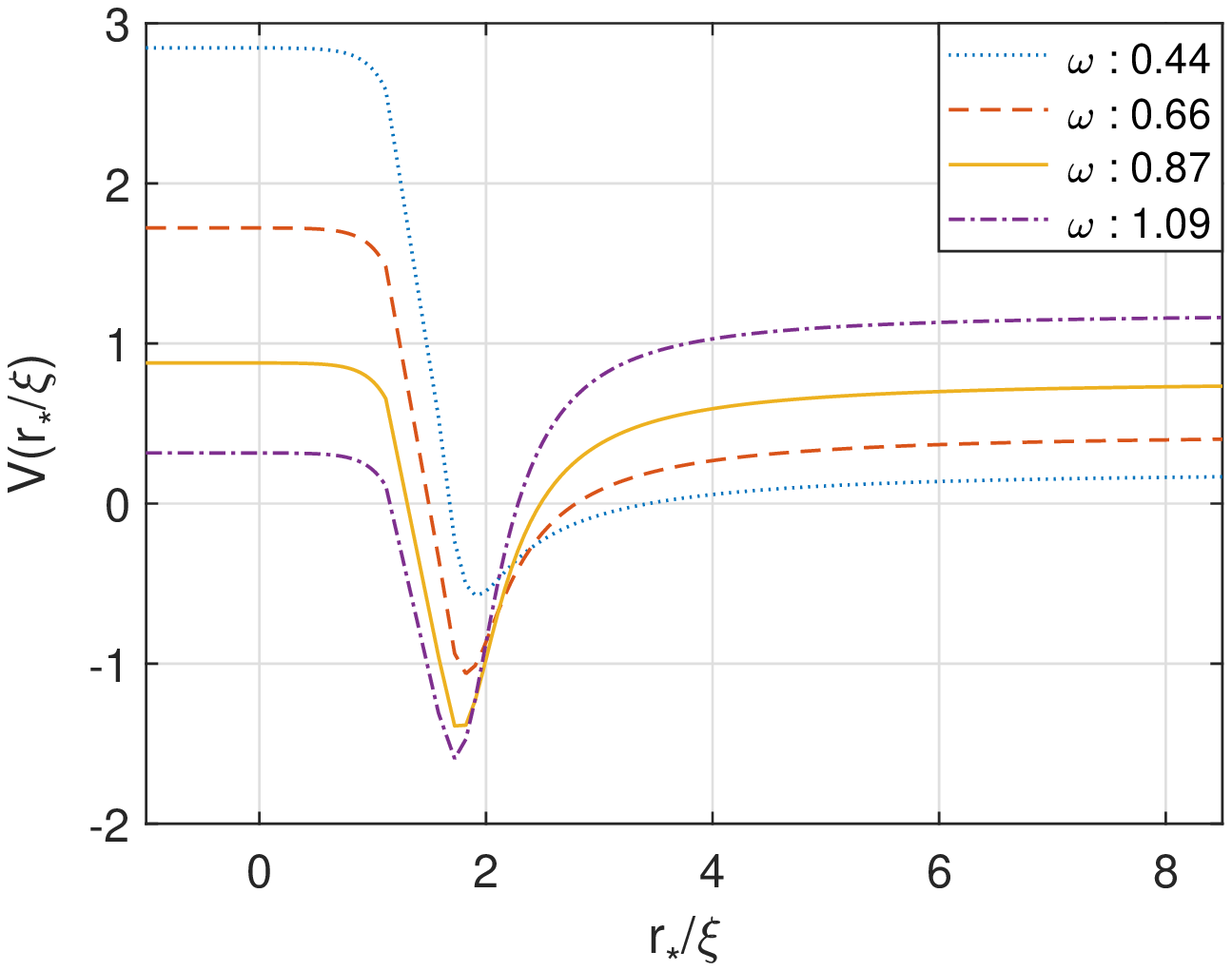}
    \end{subfigure}
    \begin{subfigure}{0.4\textwidth}
        \includegraphics[width=\textwidth]{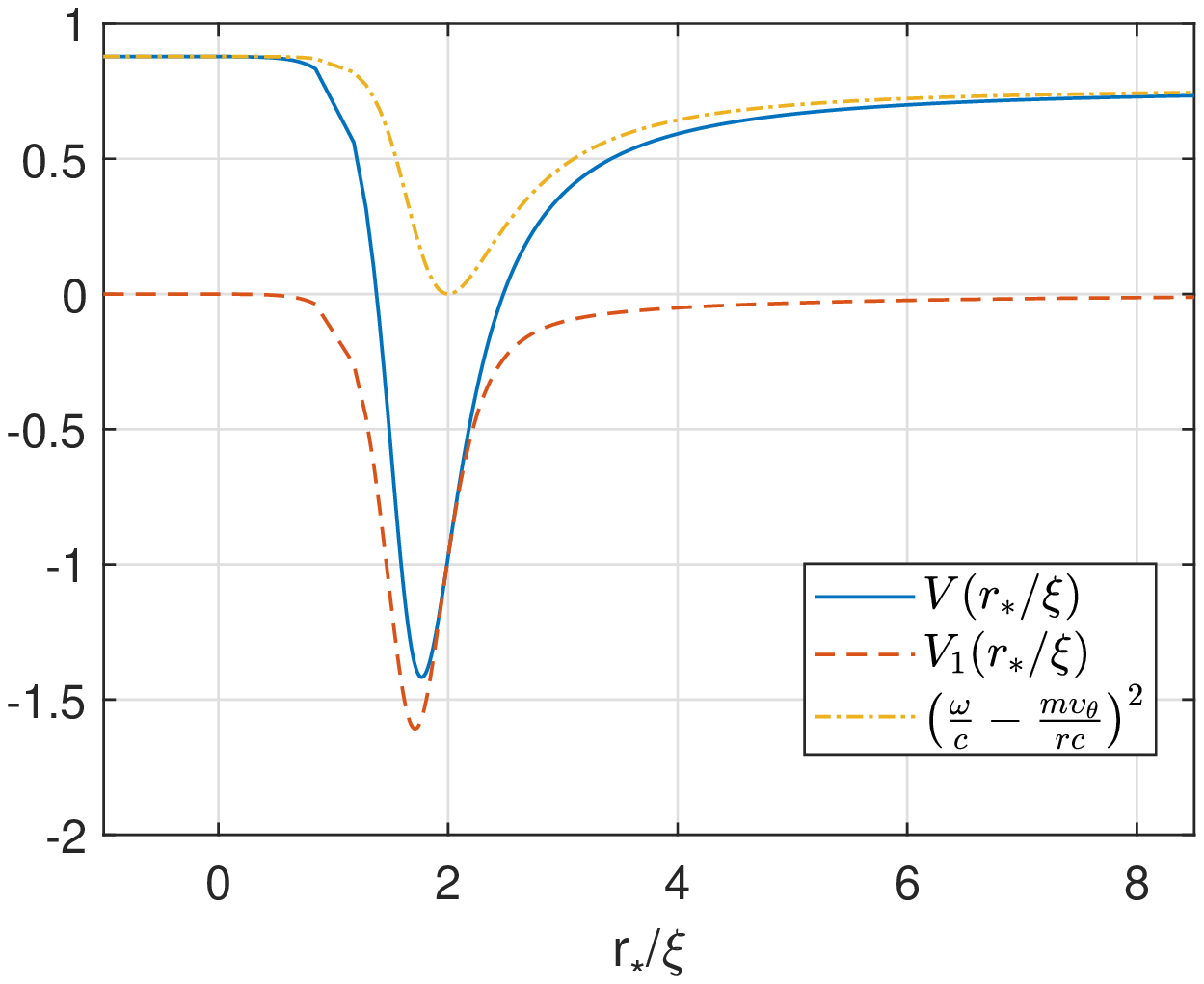}
    \end{subfigure}
    \caption{The radial behavior of $V(\tilde{r_*},\tilde{\omega})$ for $\tilde{\omega}= 0.44, 0.66, 0.87, 1.09$, and behaviors of $E(\tilde{r},\tilde{\omega})$ ,$V_1(\tilde{r})$ at fixed $\tilde{\omega}=0.87$ for $m=1$}.
		\label{fig:pot1}
\end{figure}

Figure \ref{fig:pot1}, panel (a) shows the radial behavior of $V(\tilde{r_*},\tilde{\omega})$ for $m=1$ with different values of $\tilde{\omega}$ and panel (b) shows the contributions of the terms to $V(\tilde{r_*},\tilde{\omega})$ as defined above at fixed $\tilde{\omega}=0.87$. Figure 6 shows the same quantities for $m=0$. The comparison provides features that correlate with the existence and the maximum of the superradiance: The ordering of the $\tilde{\omega}$-level curves at either end of the range are reversed for the superradiant case (Fig.5(a)) whereas they are simply shifted for the $m=0$ case. Interestingly, the maximum superradiance is achieved when the asymptotic values of $V(\tilde{r_*}\rightarrow +\infty)=V(\tilde{r_*}\rightarrow -\infty)$, i.e. the value of the energy-potential function at the event horizon matches the far-field value.  Since $V_1 (\tilde{r_*}\rightarrow +\infty)=0$ as shown in Fig.\ref{fig:pot1} (b), the maximum superradiance is in fact controlled by the $E(\tilde{r_*},\tilde{\omega})$ which makes a dip between the event horizon and the ergosphere. In contrast, in the non-superradiant $(m=0)$ case, $E(\tilde{r_*},\tilde{\omega})$ is a monotonically increasing function from its asymptote towards the event horizon.

\begin{figure}[H]
    \centering
    \begin{subfigure}{0.4\textwidth}
        \includegraphics[width=\textwidth]{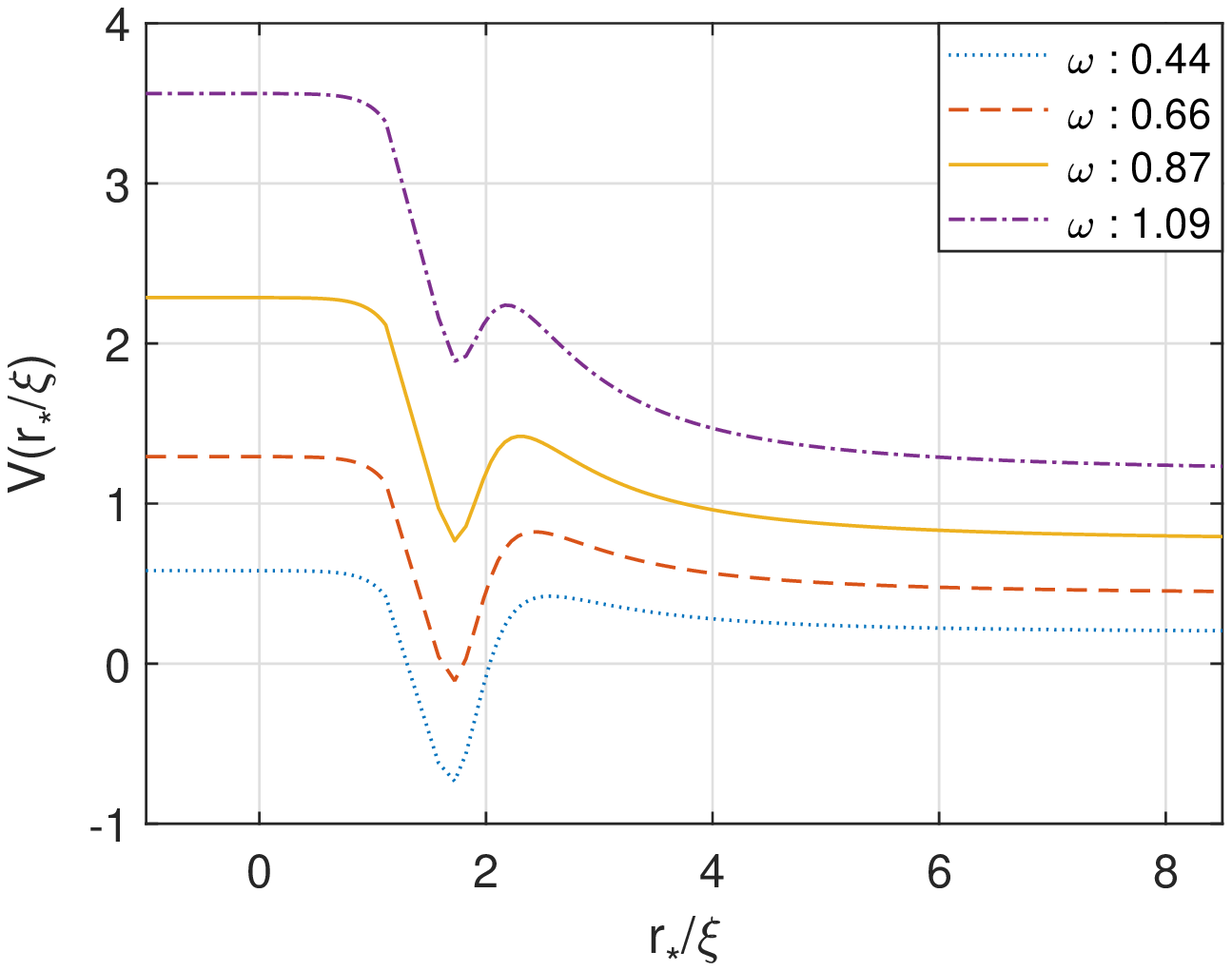}
    \end{subfigure}
    \begin{subfigure}{0.4\textwidth}
        \includegraphics[width=\textwidth]{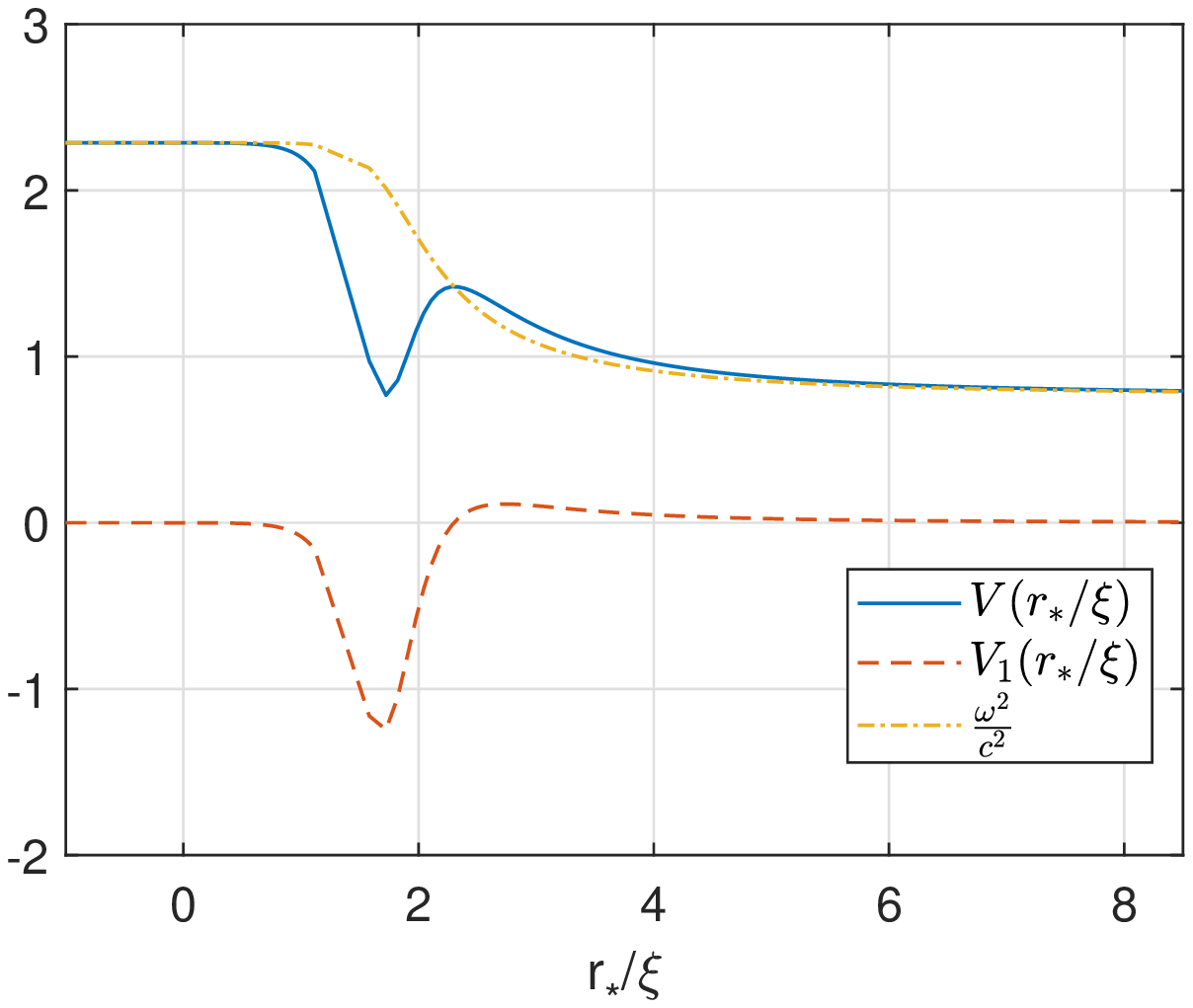}
    \end{subfigure}
    \caption{The radial behavior of $V(\tilde{r_*},\tilde{\omega})$ for $\tilde{\omega}= 0.44, 0.66, 0.87, 1.09$, and behaviors of $E(\tilde{r},\tilde{\omega})$ ,$V_1(\tilde{r})$ at fixed $\tilde{\omega}=0.87$ for $m=0$ .}
		\label{fig:pot2}
\end{figure}

\begin{figure}[H]
	\centering
		\includegraphics[width=0.6\textwidth]{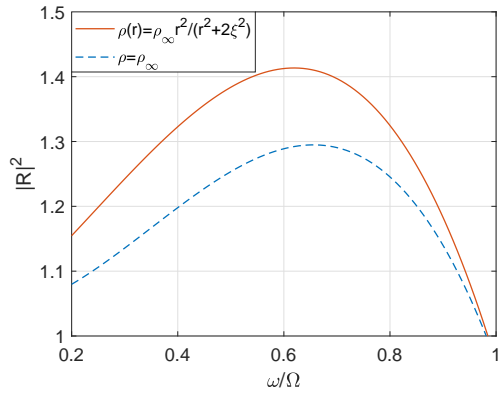}
	\caption{$\left|R\right|^2$ is calculated for superradiant case with $\tilde{\Omega}=\sqrt{2}$ in constant and non-constant density profiles.}
	\label{refcoef}
\end{figure}

Returning to the perturbative wave (the massless scalar field in Eq.\ref{eq:wavefunc}) we have
\begin{equation}
\Phi_1 = \left(\frac{r^2 + 2\xi^2}{r^3}\right)^{1/2} H(r_*)e^{i(m\theta -\omega t)e^{i\hat{h}(r)}}.
\label{eq:}
\end{equation} 
Near the event horizon and at $r \rightarrow +\infty$, the solution of  Eq.\ref{eq:H} reduces asymptotically to harmonic solutions
					 \begin{equation}
H({r_*}) = e^{i\frac{\omega}{c_{\infty}} r_*} + {\mathop{\rm \mathcal{R}e}\nolimits} ^{ - i\frac{\omega}{c_{\infty}} r_*} ,  r^* \stackrel{}{\rightarrow} +\infty
   \end{equation}
	
	 \begin{equation}
      H({r_*}) = \mathcal{T}e^{- \frac{i}{c(r_h)}\left(\omega  - \frac{m\upsilon_{\hat{\phi}}(r_h)}{r_h}\right)r^*}, r^* \stackrel{}{\rightarrow} -\infty.
      \end{equation}
Which introduces complex transmission amplitude through $r_h$ and a complex reflection amplitude at $r \rightarrow \infty$ (of an outgoing wave) for a perturbative wave of unit amplitude incident to the vortex. There are two linearly independent (complex conjugate) solutions whose Wronskian is constant everywhere. Using the rotational speed of the vortex $\Omega$, the relation between the reflection and transmission coefficients can be expressed as
			
\begin{equation}
             1 - {\left| R \right|^2} = \sqrt{3}\left( {\frac{{\omega  - m\Omega }}{\omega }} \right)\left| {{T^2}} \right|,
						\label{supercond}
     \end{equation}
Solving Eq.\ref{eq:tort} and Eq.\ref{eq:H}, the reflection coefficient $\left|\mathcal{R}\right|^2$ can be calculated through the Fourier components of the asymptotic far field solutions. Figure \ref{refcoef} shows the spectra of $\left|\mathcal{R}\right|^2$ for constant ans non-constant density approximations, which differ in all but the $\omega/\Omega \cong 1$ limit. Figure \ref{refcoef} shows the spectral comparison of the reflection coefficient calculated with constant and non-constant background density profiles, respectively. In all but the high frequency part of the spectrum, the reflection coefficient differs substantially, for which higher amplification is found for the non-constant density case.

\section{Discussion}
In this work, the amplified scattering of axisymmetric acoustic perturbations from the vortex state of a BEC are studied with a self-consistent background density profile. The density profile and a characteristic length scale (e.g. the healing length) used in the formulation inherently defines the vortex parameters (event horizon, angular speed). This is a prominent feature not available in the constant background density approximation. In fact, under constant background density, the rotational speed of the vortex stands as an independent parameter that can be changed arbitrarily.
When the frequency of the incident perturbative wave is close to that of the vortex rotational speed $\omega \approx \Omega$ the constant background density agrees well with the result of the non-constant density. At lower frequencies, the non-constant density provides higher superradiance.
A prospective study is to extend the present formulation to bound BEC subject to a confining potential, coaxial with the vortex. The "spilling" of the superradiance from the asymptotic edge of the confining potential would be an interesting feature to investigate.

\bibliographystyle{unsrt}
\bibliography{bibfile} 

\end{document}